\documentclass[10pt,prb,aps,twocolumn,showpacs,floatfix]{revtex4-1}
\usepackage{graphicx}
\usepackage{amssymb,amsmath}
\usepackage{wasysym}
\usepackage{bm}
\usepackage{pstricks}
\usepackage{epsfig}

\newcommand{\be}{\begin{eqnarray}}
\newcommand{\ee}{\end{eqnarray}}

\begin{document}

\title{Example of a first-order N\'eel to Valence-Bond-Solid transition in two-dimensions}

\author{Arnab Sen and Anders W. Sandvik}

\affiliation{Department of Physics, Boston University, 590 Commonwealth Avenue, Boston, Massachusetts 02215, USA}

\date{\today}
\begin{abstract}
We consider the $S=1/2$ Heisenberg model with nearest-neighbor interaction $J$ and an additional multi-spin interaction $Q_3$ 
on the square lattice.  The $Q_3$ term consists of three bond-singlet projectors and is chosen to favor the formation of a 
valence-bond solid (VBS) where the valence bonds (singlet pairs) form a staggered pattern. The model exhibits a quantum phase 
transition from the N\'eel state to the VBS as a function of $Q_3/J$. We study the model using quantum Monte Carlo (stochastic 
series expansion) simulations. The N\'eel--VBS transition in this case is strongly first-order in nature, in contrast to similar 
previously studied models with continuous transitions into columnar VBS states. The qualitatively different transitions 
illustrate the important role of an emerging U($1$) symmetry in the latter case, which is not possible in the present model 
due to the staggered VBS pattern (which does not allow local fluctuations necessary to rotate the local VBS order parameter).
\end{abstract}


\maketitle

\section{Introduction} 

Even though quantum phase transitions (QPTs) occur strictly at zero temperature (T), they can strongly influence the behaviour of a system 
at $T>0$ in their vicinity.\cite{Subirbook} Exotic QPTs that do not have any analogs in classical critical phenomena may thus leave widely
observable signatures.  A very interesting example of such a QPT was proposed by Senthil~{\it et al.},\cite{Senthil_etal} who reconsidered
the transition between N\'eel and valence-bond-solid (VBS) phases in two-dimensional (2D) square-lattice $S=1/2$ antiferromagnetic spin 
systems.\cite{read,murthy}  They argued that subtle Berry phase effects can lead to an unconventional ``deconfined'' quantum-critical 
(DQC) transition that is generically continuous, while a standard Landau-Ginzburg-Wilson analysis would instead predict a first-order 
transition (except at fine-tuned multi-critical points), since the two phases break unrelated symmetries. At the DQC point, new fractionalized 
degrees of freedom, $S=1/2$ spinons, become liberated,\cite{Senthil_etal} while they are confined in the conventional phases on either 
side of the transition. 

An important question raised by the field-theoretical work on DQC points\cite{Senthil_etal} is whether there are microscopic models (hamiltonians) 
where the existence of this unconventional quantum-criticality can be explicitly demonstrated. Such models, apart from allowing unbiased tests of the 
validity of the concept, can lead to further theoretical developments. This is also important in view of the work reported by 
Kuklov {\it et al.} in Refs~\onlinecite{Kuklov1,Kuklov2}, where an action claimed to realize the low-energy physics of the non-compact (NC) 
CP$^{1}$ field-theory, argued by Senthil {\it et al.} to describe the DQC point,\cite{Senthil_etal} was studied using Monte Carlo simulations of
3D classical current models with easy-plane and $SU(2)$ symmetry. The transitions in both cases were argued to be generically  weakly 
first-order, in contradiction with the proposal of Ref.~\onlinecite{Senthil_etal}. Other, similar studies have given different results, 
however.\cite{motrunich,takashima} A full understanding of the NCCP$^{1}$ theory and its ability to capture the physics of the N\'eel--VBS transition 
in 2D antiferromagnets is still lacking. 
 
One approach to the DQC problem is to identify and study microscopic quantum spin hamiltonians exhibiting the N\'eel--VBS transition. In a recently 
proposed class of $S=1/2$ ``J-Q'' models,\cite{Anders_jq,Louetal} a geometrically unfrustrated multi-spin interaction of strength $Q$ (consisting of two or 
more bond-singlet projectors) was introduced which competes with the Heisenberg interaction $J$. Quantum Monte Carlo (QMC) calculations in this case 
are free from the sign problems \cite{henelius} hampering studies of frustrated systems (which also exhibit N\'eel--VBS transitions \cite{jjmodel}). 
Calculations in the ground state \cite{Anders_jq,Louetal} and at finite temperature,\cite{melkokaul} have demonstrated behaviors consistent with a 
continuous N\'eel-VBS transition of the DQC type. In particular, the dynamic exponent $z=1$ and the anomalous dimension associated with the spin 
correlations is relatively large; $\eta\approx 0.35$. There does, however, appear to be significant corrections to scaling at the critical point, 
possibly of the logarithmic type.\cite{Anders_jqlog} Such corrections, which have also been observed by studying the response of the spin texture to 
impurities,\cite{kedar} raise questions on the presence of marginally irrelevant operators (or, possibly, conventional operators with scaling dimension 
very close to $0$) in corresponding field-theories. It has also been claimed that observed scaling corrections indicate an eventual first-order 
transition, \cite{Kuklov2} which, however, is unlikely, considering the fact that the scaling corrections are well behaved 
over a large range of system sizes.\cite{Anders_jqlog} Some scaling corrections are always expected, and it can be 
noted that logarithmic corrections have  been found in related gauge theories with fermions.\cite{kim}

\begin{figure}
\centerline{\includegraphics[width=6.2cm, clip]{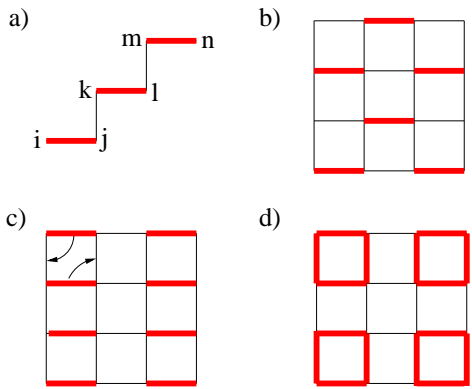}}
\caption{(Color online) a) The interaction term $Q_3$ involving three bond-singlet projection operators (shown with thicker, red lines) on the square 
lattice. All terms related by lattice translations and rotations of the shown instance of a product of singlet projectors are included in the 
Hamiltonian. Examples of VBS patterns: b) staggered, c) columnar, and d) plaquette. The singlets preferentially form on the thicker (red) bonds. 
The arrows in c) indicate how a local resonance of a pair of bonds between horizontal and vertical orientation corresponds to a plaquette
in d). Such resonances correspond to fluctuations of the VBS angle, which is $\phi=n\pi/2$ ($n=0,1,2,3$) for the columnar state and $\phi n\pi/2+\pi/4$ 
for the plaquette state. In the DQC scenario \cite{Senthil_etal} they lead to an emergent U$(1)$ symmetry, i.e., a continuous circular-symmetric $\phi$ 
as the transition into the N\'eel state is approached. The $Q_3$ term favors the staggered-type VBS b), where such angular fluctuations
should always be small.}
\label{fig1}
\end{figure}

In light of the controversy regarding the nature of the N\'eel--VBS transition, and in order to gain further insights into the unusual physics of DQC 
points, it is useful to also study similar transitions that are definitely of first order. This is the topic of the present paper. Ideally, one would like 
to have a model in which the transition could be tuned from strongly to weakly first-order, through a tricritical point into a continuous transition. 
No such tunable model is known so far, however. In this work, we provide an example of a N\'eel-VBS transition on the 2D square lattice that is strongly 
first order. We introduce a J-Q$_{\rm 3}$ model in which the Q$_{\rm 3}$ interaction consists of three singlet projectors arranged in a staggered fashion 
on the square lattice, as illustrated in Fig.~\ref{fig1}(a). This interaction induces a staggered VBS pattern in which resonating valence bonds 
leading locally to a different (degenerate) VBS pattern are not favored. If such fluctuations are present they can effectively rotate the coarse-grained 
angle of the VBS order-parameter (as explained further in the caption of Fig.~\ref{fig1}), which has been explicitly observed in the J-Q models studied 
previously.\cite{Anders_jq,jiang,Louetal} In the DQC scenario, they are directly responsible for the emergent U$(1)$ symmetry of the VBS order 
parameter in the neighborhood of the transition.\cite{Senthil_etal} The absence of this feature in the model studied here brings it clearly outside 
the framework of DQC points, and a numerical confirmation of a different type of transition is then, indirectly, an additional piece of evidence in favor 
of a consistent DQC scenario in which emergent U(1) symmetry and spinon deconfinement should go hand-in-hand with a continuous transition. 

We here use the stochastic series expansion (SSE) QMC method with operator-loop updates \cite{sse} to study the nature of the N\'eel--VBS transition 
in the staggered J-Q$_{\rm 3}$ model. We perform simulations at a fixed aspect ratio of inverse temperature $\beta J=L$, as done previously for the 
standard J-Q$_{\rm 2}$ model in Refs.~\onlinecite{melkokaul,jiang,Anders_jqlog}. We study the finite-size scaling properties of various physical quantities 
and contrast them with what is observed at the previously studied putative continuous DQCs.

The rest of the paper is organized in following way: In Sec II, we define the model more precisely and present the results for the staggered magnetization, 
the corresponding Binder cumulant, the spin stiffness, and the VBS order parameter. We also consider the probability distribution of the VBS order 
parameter and use it to explicitly demonstrate phase coexistence. In Sec III, we determine the location of the critical point 
by using the crossing of the energies of the N\'eel and the VBS phases in the meta-stable region near the transition. We state our conclusions and 
discuss future prospects in Sec IV. 

\section{Model and order parameters} 

We consider the following Hamiltonian
\be 
H=J\sum_{\langle ij \rangle}\mathbf{S}_i \cdot \mathbf{S}_j - Q_3 \sum_{\langle ijklmn \rangle}C_{ij}C_{kl}C_{mn},
\label{hamiltonian}
\ee
where ${\bf{S}}_i$ refers to a $S=1/2$ spin at site $i$ on the 2D square lattice and $C_{ij}$ denotes the singlet pair projection operator,
\be
C_{ij}=\hbox{$\frac{1}{4}$}-\mathbf{S}_i \cdot \mathbf{S}_j, 
\ee
between two nearest neighbors $i$ and $j$. The $Q_3$ term (where, in the notation of Ref.~\onlinecite{Louetal}, the subscript on $Q$ refers to the number of singlet 
projectors in the product) is chosen in the particular manner illustrated in Fig~\ref{fig1}(a), to favor the formation of the kind of staggered VBS
illustrated in Fig~\ref{fig1}(b). Like the columnar and plaquette VBS, the broken symmetry of the staggered VBS is $Z_4$. However, this type of VBS is very 
different from its columnar or plaquette counterparts, since no local ring-exchange of singlets on closed loops [e.g., as illustrated in Fig~\ref{fig1}(c) for 
a simple two-bond resonance] is possible in the {\it ideal} staggered VBS. This makes it highly unlikely for the existing fluctuations of this kind of VBS
to be associated with an emergent $U(1)$ symmetry, which is a key characteristic of the DQC transition.\cite{Senthil_etal} We will confirm this with
simulation results below.

We have also studied an interaction similar to the six-spin $Q_3$ term but with only two singlet projectors, on two pairs of sites separated by one 
lattice spacing and shifted one step with respect to each other as in Fig.~\ref{fig1}(a). This interaction is not sufficient for destroying the N\'eel order, 
however, unlike the original J-Q model with the two singlet projectors inside $2\times 2$ plaquettes. In the latter case the resulting VBS in the extreme 
case of $J=0$ is also quite weak,\cite{Anders_jq} while adding one more singlet projector (with the sets of three projectors arranged in columns) gives 
a much more robust VBS order.\cite{Louetal}

To study the N\'eel-VBS phase transition in the staggered J-Q$_{\rm 3}$ model (\ref{hamiltonian}),
we measure quantities that are sensitive to the N\'eel order and the VBS order respectively. At a continuous
quantum phase transition, these quantities should scale with the system size $L$ according to non-trivial critical exponents, while at a first-order transition 
one would expect very different exponents related to the dimensionality of the system as well particular signatures of coexisting phases at the transition point.
These signatures should apply when the linear dimension of the system $L \agt \xi$, where $\xi$ is the finite correlation length at the transition.

\subsection{N\'eel order} 

\begin{figure}
\centerline{\includegraphics[width=\hsize, clip]{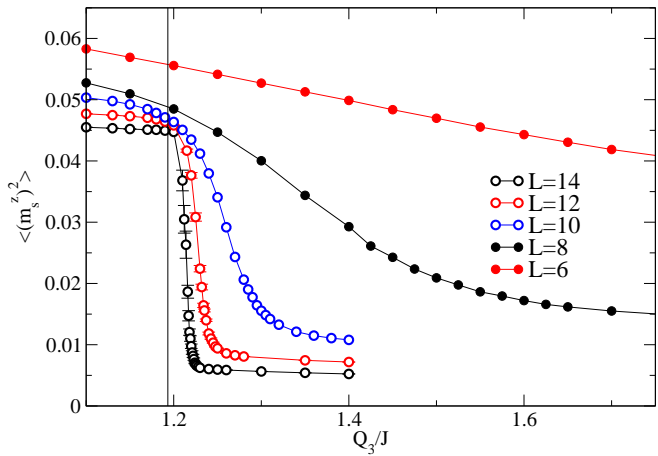}}
\caption{(Color online) The squared staggered magnetization $\langle (m_s^z)^2 \rangle$ shown for different system sizes at inverse temperature 
$\beta J = L$. The vertical line at $(Q_3/J)_c=1.1933$ is the estimated $L\to \infty$ transition point from crossings of meta-stable energy branches 
(Fig~\ref{fig7}).}
\label{fig2}
\end{figure}

The magnetically ordered N\'eel phase breaks the SU($2$) rotational symmetry of the interaction hamiltonian $H$ and can be characterized by measuring 
$\langle (m^z_s)^2 \rangle$, where $m^z_s$ denotes the $z$-component of staggered magnetization of the system: 
\be 
m^z_s=\frac{1}{N} \sum_\mathbf{r} S^z(\mathbf{r})\cos(\mathbf{Q} \cdot \mathbf{r})
\ee
with ${\bf{Q}}=(\pi,\pi)$ the wave-vector corresponding to the N\'eel phase and $N=L^2$. This quantity is diagonal in the $S^z$ basis used and can be easily 
measured in the SSE simulations. We measure the squared quantity $\langle (m_s^z)^2 \rangle$, which, due to the spin-rotational symmetry of the hamiltonian, 
is $1/3$ of the full squared staggered magnetization $\langle m_s^2 \rangle$. We show the data for different system sizes at $\beta J=L$ near the phase transition 
in Fig~\ref{fig2}. As the system size is increased, we observe a jump developing in $\langle (m^z_s)^2 \rangle$ that becomes more abrupt and rapidly approaches 
the infinite-volume estimate of the critical point $(Q_3/J)_c=1.1933(1)$ for this model (indicated by the vertical line in Fig~\ref{fig2} and other figures). 
The value of $(Q_3/J)_c$ was obtained from the crossing of metastable energies of the N\'eel and VBS phases of larger systems; see Fig~\ref{fig7} and later 
discussion in Sec.~\ref{transitionpoint}. This kind of behaviour of the N\'eel order-parameter is already very suggestive of a first-order transition. Data for 
systems larger than $L=14$ are not shown here because of the extremely long tunneling times between the co-existing (as we will show below in Sec,~\ref{vbsorder}) 
N\'eel and VBS phases for such sizes in our simulations close to the phase transition, which makes it very difficult to obtain reliable expectation values. 

\begin{figure}
\centerline{\includegraphics[width=\hsize, clip]{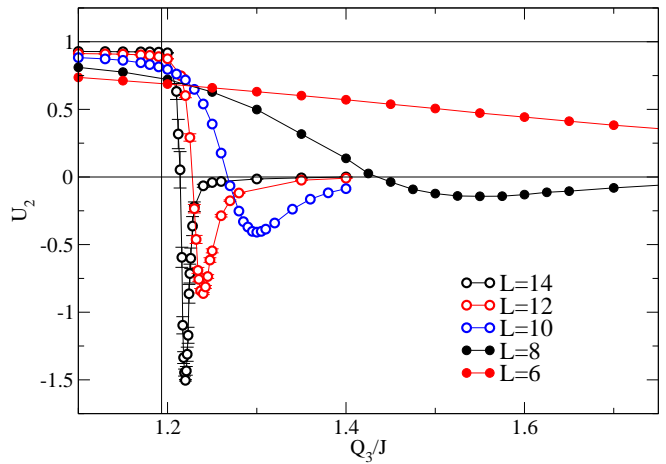}}
\caption{(Color online) The Binder cummulant of the staggered magnetization shown for different system sizes at inverse temperature $\beta J=L$. 
Note that the minimum of the Binder cummulant is negative for $L \ge 8$ and diverges to $-\infty$ as $L \rightarrow \infty$ based on these sizes.}
\label{fig3}
\end{figure}

A quantity that is very useful for distinguishing between first-order and continuous phase transitions is the Binder cummulant $U_2$, defined for
an O$(3)$ order parameter as \cite{Binder}
\be
U_2=\frac{5}{2}\left( 1- \frac{\langle (m_s^z)^4 \rangle}{3 \langle (m_s^z)^2 \rangle^2}\right).
\ee
With the factors used here, $U_2 \rightarrow 1$ in the N\'eel phase and $U_2 \rightarrow 0$ in the magnetically disordered phase (VBS in this case) when 
$L \rightarrow \infty$. For a continuous phase transition, the $U_2$ curves for different system sizes intersect at the critical point (for sufficiently 
large L) and the value of $U_2$ at the intercept normally lies in the interval $(0,1)$.\cite{Binder} This property of the Binder cummulant is often used to accurately 
determine the location of the critical point for continuous phase transitions. However, for a first order transition, the Binder cummulant behaves in a 
completely different manner, that was explained phenomenologically for classical transitions by Vollmayr {\it et al.}\cite{Vollmayr} For systems exceeding 
a certain length $L_{\rm min} \sim \xi$, the curves show a minimum which becomes more pronounced as the system size increases. The minimum value of 
$U_2 \rightarrow -\infty$ as $L \rightarrow \infty$ because of phase co-existence, and the position of the minimum approaches the transition point in the 
thermodynamic limit. This behaviour has been observed in previous studies of classical first order phase transitions [for example, see Refs~\onlinecite{Vollmayr,Cepas}]. 
Indeed, the Binder cummulant for the J-Q$_{\rm 3}$ model, graphed in Fig~\ref{fig3}, behaves in a similar manner and strongly points to a first order phase transition. 
The negative minimum is present in the $U_2$ curves for $L=6$ and above and becomes deeper and sharper as $L$ increases. Its location approaches the estimated
$(Q_3/J)_c$. The fact that a minimum in $U_2$ is not at all present for $L=4$ and is barely negative for $L=6$ allows us to estimate that the typical length-scale
(the spin correlation length) at the first-order transition is approximately in the range $\xi = 4 \sim 6$. We also measure the second moment spin correlation length $\xi_a$ defined as 
\be 
\xi_a = \frac{L}{2 \pi}\sqrt{\frac{S(\pi,\pi)}{S(\pi+\frac{2\pi}{L},\pi)}-1}
\ee 
where $S({\bf{q}})$ refers to the spin structure factor at the corresponding wavevector ${\bf{q}}$. We obtain $\xi_a \sim 2$ close to the critical point in the magnetically disordered VBS phase. However, when the spin correlation length is small, $\xi_a$ can differ from the {\it true} correlation length $\xi$ based on the asymptotic decay of the spin correlations in real space (which is difficult to extract reliably).

\begin{figure}
\centerline{\includegraphics[width=\hsize, clip]{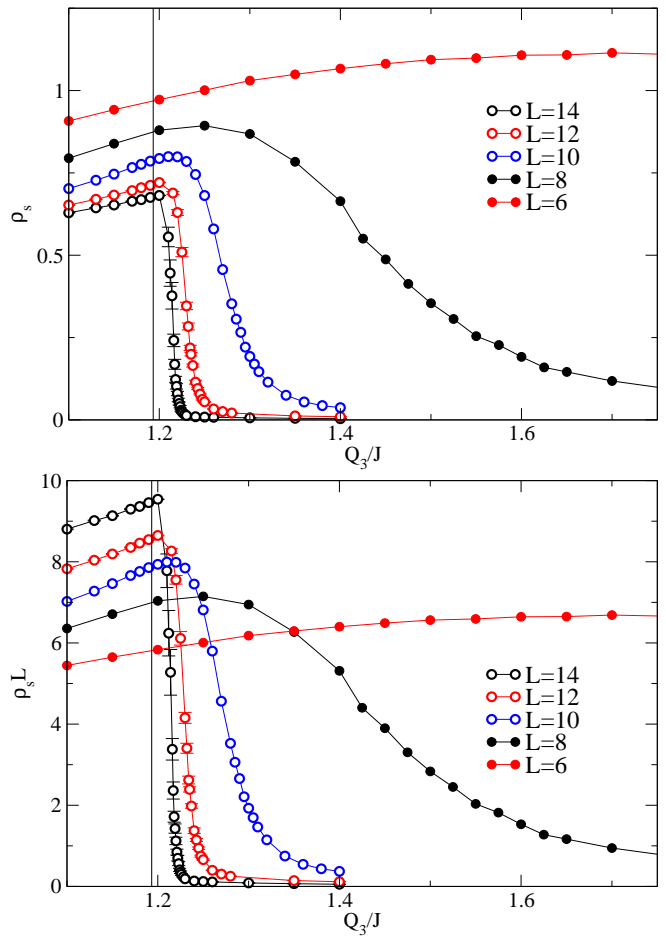}}
\caption{(Color online) The spin stiffness $\rho_s$ (top panel) and $\rho_sL$ (bottom panel) for different system sizes.}
\label{fig4}
\end{figure}

An important quantity characterizing the N\'eel phase is the spin stiffness $\rho_s$; the second derivative of the ground state energy $E(\phi)$ (per spin) 
in the presence of a twisted boundary condition (with rotation of a boundary row or column by an angle $\phi$);
\begin{equation}
\rho_s=\frac{\partial^2E(\phi)}{\partial \phi^2},
\end{equation}
This quantity is obtained in the SSE simulations in the standard way using the fluctuation of the winding numbers.\cite{pollock,sandvik97} Again, the 
finite-size behaviour should be very different for a continous $z=1$ phase transition and for a first order transition. In the former case, 
$\rho_s \sim 1/L $ for large enough system sizes at the critical coupling. Thus, $\rho_s L$ curves for different system sizes intersect at the critical coupling 
for a $z=1$ continous critical point, given $L$ is sufficiently large. However, at a first order transition, since there is phase coexistence, $\rho_s$ is 
finite (non-zero) at the transition point when $L \rightarrow \infty$ and thus $\rho_s L \rightarrow \infty$ as the system size is increased and no unique crossing 
point is obtained. Fig~\ref{fig4} shows the behavior of both $\rho_s$ and $\rho_sL$. Intersection between $\rho_sL$ curves for different $L$ can indeed be
seen close to the transition point, but the value at the intersection points (for, e.g., $L$ and $L+2$) increases with $L$. The $\rho_s$ curves become 
sharper and approach a step function with a discontinuity at $(Q_3/J)_c$ as $L$ increases. Interestingly, the spin stiffness $\rho_s$ actually increases on approaching the transition from the N\'eel phase before sharply dropping off in the VBS phase.

\subsection{VBS order} 
\label{vbsorder}

\begin{figure}
\centerline{\includegraphics[width=\hsize, clip]{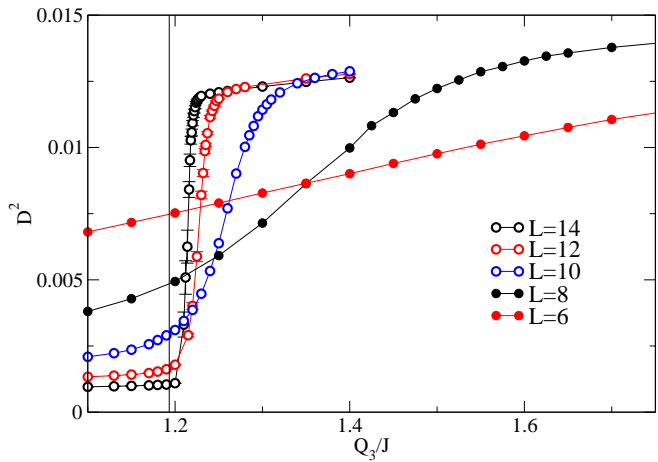}}
\caption{(Color online) Size dependence of the VBS order parameter $D^2$ for different system sizes.}
\label{fig5}
\end{figure}

To provide further evidence for the first order nature of the transition, and to obtain a bench-mark for the finite-size scaling,
it is useful to also consider observables sensitive to the VBS order. We define the following order parameters for that purpose:
\be 
D_x &=& \frac{1}{N}\sum_\mathbf{r} S^z(\mathbf{r})S^z(\mathbf{r}+\hat{\mathbf{x}})\cos(\mathbf{Q} \cdot \mathbf{r})  \\
D_y &=& \frac{1}{N}\sum_\mathbf{r} S^z(\mathbf{r})S^z(\mathbf{r}+\hat{\mathbf{y}})\cos(\mathbf{Q} \cdot \mathbf{r})  \\
D^2 &=& \langle D_x^2+D_y^2 \rangle
\ee
where again $\mathbf{Q}=(\pi,\pi)$. The squared order parameter $D^2$ is zero in the Neel phase and becomes non-zero in the VBS phase in the thermodynamic limit. 
In Fig~\ref{fig5}, we show the behaviour of $D^2$ near the transition. The jump in $D^2$ gets sharper with increasing system size, similar to the staggered 
magnetization in Fig~\ref{fig2}, and its location approaches $(Q_3/J)_c$. Thus, the solid order parameter also shows a first order transition. Note that
the order parameter we use here is not spin-rotationally invariant. One can also define $D_x$ and $D_y$ using rotationally invariant bond operators
such as $\mathbf{S}(\mathbf{r})\cdot \mathbf{S}(\mathbf{r}+\hat{\mathbf{x}})$. There may be differences in the finite-size scaling properties,\cite{melkokaul} but for our purposes here the rotational
invariance is not necessary (and the $z$-component definition of $D^2$ is easier to measure with the SSE method).

A strong signature of a first order transition is phase co-existence at the transition point, unlike in the case of a continuous transition. We already showed
evidence of co-existence of a N\'eel and non-magnetic phase in the form of a negative Binder cumulant of the staggered magnetization in Fig.~\ref{fig2}. The 
coexistence can be observed in a more direct way using the VBS order parameter. Consider the joint probability distribution function $P(D_x,D_y)$. In the Neel phase, 
this function is peaked at $(0,0)$. In the VBS phase, $P(D_x,D_y)$ is peaked at $(0,\pm D)$ and $(\pm D,0)$, where $D$ is finite, reflecting the $Z_4$ degeneracy 
of the VBS state in a finite lattice. Note that $(D_x,D_y)$ is here defined as a single point obtained on the basis of an equal-time simultaneous measurement of
$D_x$ and $D_y$, i.e., these operators are not averaged over the imaginary-time dimension in the simulations. The full distribution can still of course
be accumulated over several imaginary times.

\begin{figure}
\centerline{\includegraphics[width=8cm, clip]{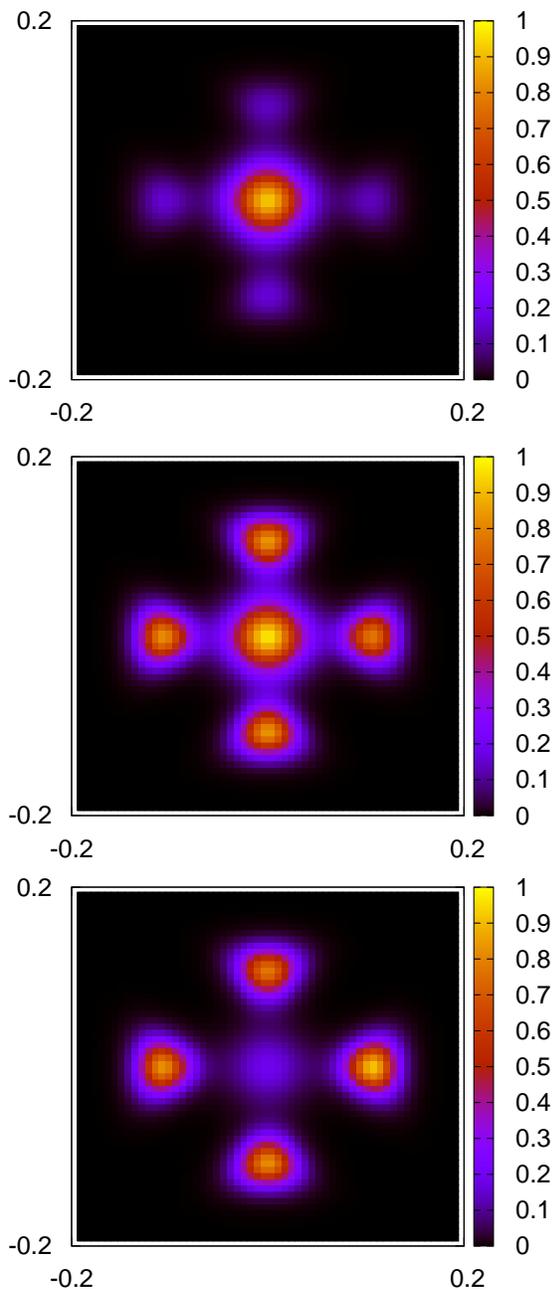}}
\caption{(Color online) The probability density $P(D_x,D_y)$ shown for $L=12$ at $Q_3/J=1.22,1.23,1.24$ and $\beta J=12$. The maxima present both at $(0,0)$ and 
$(\pm D,0),(0,\pm D)$ at $Q_3/J=1.23$ show phase coexistence of the Neel and VBS orders.}
\label{fig6}
\end{figure}

In Fig~\ref{fig6}, we show $P(D_x,D_y)$ for $L=12$ and $Q_3/J=1.22,1.23,1.24$ for $\beta J=L$. The coexistence of the Neel and the VBS phase is evident from the presence of 
peaks at both $(0,0)$ and $(\pm D,0), (0,\pm D)$ at $Q_3/J=1.23$, while at $Q_3/J=1.22$ ($Q_3/J=1.24$), the N\'eel (VBS) phase dominates. Also note the absence of any $U(1)$ ring like feature in the distribution shown in Fig~\ref{fig6}. 
This should be contrasted with similar measures of the distribution function in the J-Q models with columnar VBS states close to the critical point, where the 
enlarged $U(1)$ symmetry is very evident.\cite{Anders_jq, jiang,Louetal} In the original J-Q model with two singlet projectors, the VBS does not seem to get pinned 
to the four $Z_4$ symmetric angles even at $J=0$ for the system sizes accessible.\cite{Anders_jq} In the modified J-Q model with three singlet projectors forming
columns,\cite{Louetal} the change in the shape of the VBS order parameter distribution from $U(1)$ close to the transition to $Z_4$ deep in the VBS phase can 
be clearly observed, however. The enlarged $U(1)$ symmetry arises in DQC theory due to the (dangerously) irrelevant $Z_4$ symmetry-breaking term at the critical 
point.\cite{Senthil_etal} Away from the critical point, the symmetry is only approximate but {\it large} system sizes are needed to observe that (i.e., $L$
has to exceed the length scale $\Lambda$  which is larger than the standard correlation length and determined by the dangerously irrelevant operator). 

Within the DQC framework, the approximate $U(1)$ symmetry near the critical point can be thought of in the following manner:\cite{Senthil_etal} the dangerously 
irrelevant $Z_4$ perturbation only produces a small energy difference between the columnar and plaquette VBS which vanishes at the critical point, and gives 
rise to a corresponding large length scale slightly away from it within which the magnitude of the VBS order parameter is formed, but its angle (which can
be defined exactly as we did above in terms of $D_x$ and $D_y$) is not pinned in any particular direction. This feature has also been observed in $U(1)$ symmetric 
spin models where either a very {\it weak} first order transition~\cite{weak_fo} or transition with {\it unusual} finite-size scaling~\cite{JKmodel} takes place. 
However, when the VBS is staggered, there is no competing solid that is energetically close because of the absence of local ring-exchange moves of the singlets 
(or dimers) in the ideal staggered solid. This does not allow the emergence of an approximate $U(1)$ symmetry near the transition, for which local fluctuations 
of the VBS order parameter are necessary, and puts it outside the framework of DQC points even though the phases have the same broken symmetries. Note that the 
value of $D^2$ in the VBS phase after the discontinuous jump (Fig~\ref{fig5}) is {\it close} ($\sim 74\%$) to that of an ideal staggered solid ($D^2=0.015625$), which
motivates us to classify it as a strongly first-order transition. 

We should point out here that the order parameter distribution $P(m_s^z)$ of the N\'eel order parameter does not show a clear peak structure at coexistence, because 
of the spin-rotational averaging when measuring just one of the three components of the staggered magnetization. The distribution is not sharply peaked in the N\'eel
state. In principle one could measure deviations from the rotationally averaged distribution expected \cite{gerber} for a single phase, but this signal is not as 
clear as the one seen above for the VBS order parameter. In the Binder cumulant this problem is avoided, because only even powers of $m_s$ are used and they
can be trivially related to the corresponding powers of $m_s^z$. In principle one could also measure the $x$ and $y$ components of the staggered structure factor,
and compute the distribution $P(|m_s|)$, but since these are off-diagonal operators in the basis used there are ambiguities in how to define $m^x_s$ and $m^y_s$ 
for a given configuration in which $m^s_z$ is also measured.

\section{Determination of the transition point} 
\label{transitionpoint}

The location of a phase transition in the thermodynamic limit may be determined by using finite size scaling arguments. For first-order transitions at 
finite temperature, the peak position of the specific heat $C_v$, susceptibility $\chi$ and the minimum of the Binder cummulant of the order parameter 
approach~\cite{Fisher} the thermodynamic transition temperature $T_c (L \rightarrow \infty)$ as $L^{-d}$, where $d$ is the dimensionality of the system. 
However, it is known from numerical studies of classical phase transitions (e.g in $q>4$ Potts models in 2D) that large system sizes may be needed to 
observe this {\it simple} scaling of $T_c(L)$. For example, in Ref.~\onlinecite{Lee}, both the $q=8$ and $q=10$ Potts model (both of which exhibit 
first-order transitions) were studied for $L \leq 50$, and strong deviations from the $L^{-d}$ scaling was observed for the $q=8$ case at these system 
sizes. Continentino and Ferreira~\cite{Continentino} have proposed that for a quantum phase transition, similar definitions of $g_c(L)$ will approach 
$g_c(L \rightarrow \infty)$ as $L^{-(d+z)}$, where $g$ is the (non-thermal) parameter driving the quantum phase transition and $z$ is a dynamic exponent (this reflects the equivalence of the $d$-dimensional quantum model at $T=0$ to a $(d+z)$-dimensional classical model at finite temperature). In our problem, the presence of N\'eel order at the critical point and using the geometry with $\beta J=L$ imply that finite-size scaling is governed by low-energy linearly dispersing spin waves and therefore $z=1$. The location of the minimum of the Binder cummulant $U_2$ can be taken as a possible definition for such a psuedo-critical coupling $g_c(L)$. With this definition, we have not been able to check 
the above mentioned scaling for our model explicitly because of limited system sizes, i.e., the data do not follow pure power laws. To obtain reliable 
data for bigger system sizes near the critical point, where the tunneling time between the competing phases becomes very large, will possibly require 
extended ensemble methods (which have been developed for quantum system in Refs.~\onlinecite{sengupta,troyer,trebst}), which we leave for future work. 

\begin{figure}
\centerline{\includegraphics[width=\hsize, clip]{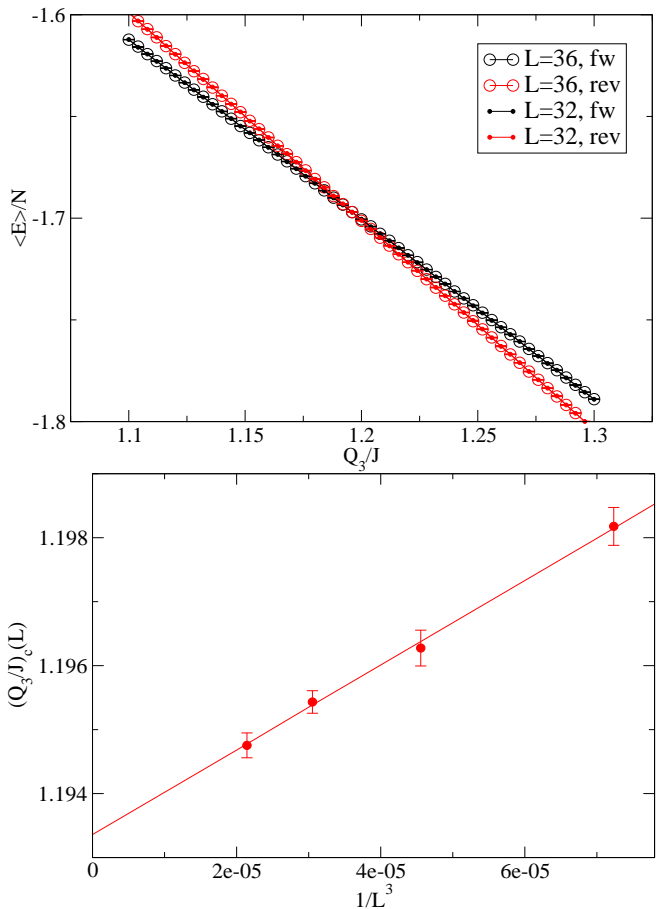}}
\caption{(Color online) Critical coupling $(Q_3/J)_c(L)$ extracted from the crossing of the energy branches (top panel) of the Neel phase (labeled as fw) and the VBS phase 
(labeled as rev) for a system of size $L=36$ (open symbols) at $\beta J=36$ and $L=32$ at $\beta J=32$ (closed 
symbols). The bottom panel shows a fit to the form $(Q_3/J)_c(L)=(Q_3/J)_c + A/L^3$ to extract the thermodynamic value of $(Q_3/J)_c=1.1933(1)$. }
\label{fig7}
\end{figure}

Interestingly, the long tunneling times associated with strong first order transitions can actually be used to locate the value of the critical point 
accurately. This can be done by determining the average energy of the two different phases~\cite{roger} in the following manner:  To determine the energy 
of the Neel phase, we start the simulation at a value of $Q_3/J$ inside that phase and then keep increasing $Q_3/J$ to go to the VBS phase, while measuring 
the average energy at each such $Q_3/J$. In this sequence, we always use the last configuration generated in the run at a particular $Q_3/J$ as the starting 
seed for the next value of $Q_3/J$. Given the system is sufficiently big so that the tunneling time is large compared to the simulation time, this procedure 
ensures that the system stays in the {\it metastable} N\'eel phase close to the critical point even when it has been crossed. We then repeat this process 
in the reverse direction starting from a value of $Q_3/J$ inside the VBS phase and decreasing the value of $Q_3/J$. The crossing point of the two energy 
branches then gives the location of a psuedo-critical coupling $(Q_3/J)_c(L)$ at that system size. In Fig~\ref{fig7} (top panel), we show the results of this procedure for $L=32$ and $L=36$. The 
curves can be fitted very well to quadratic polynomials (the small quadratic part is added to improve $\chi^2$/ dof) and the crossing point can 
be accurately determined. These crossing points are indeed consistent with a scaling of the form $(Q_3/J)_c(L)=(Q_3/J)_c-A/L^3$ for the larger system sizes studied ($L=24-36$) (see Fig~\ref{fig7} (bottom panel)) assuming that $z=1$. We obtain $(Q_3/J)_c=1.1933(1)$ using this extrapolation. 

 \section{Conclusion} 

In conclusion, we have considered a modified ``J-Q$_3$'' type Hamiltonian, where the $Q_3$ term is chosen to stabilize a staggered VBS solid. We demonstrate the 
presence of a strong first-order quantum critical point that seperates the N\'eel and the VBS phases. In the VBS phase, there is no approximate $U(1)$ symmetry and 
associated ring-like features in the probability distribution of the VBS order parameter close to the critical point. We also use the crossing of the energy branches 
of the two phases for larger system sizes to estimate the thermodynamic value of the critical point. We observe that these crossing points (for $L=24-36$) indeed converge to the thermodynamic value of the critical point as $1/L^3$, expected from finite-size scaling for first order transitions with $\beta \sim L$. It will be interesting to test similar predictions of finite-size 
scaling theory of first-order quantum phase transitions,~\cite{Continentino} which should depend solely on the dimensionality of the system $d$ and the dynamical 
exponent $z$, by measuring the order parameters and the Binder cummulant using quantum Monte Carlo simulations in combination with extended ensemble 
methods~\cite{sengupta,troyer,trebst} for bigger systems that may avoid the large tunneling times between the two co-existing phases near the transition. It will 
also be interesting to see whether there is a finite-temperature crossover to a continuous phase transition between a VBS and a disordered phase. Given the strong 
first order nature of the transition, it will likely survive at low $T$ as well. Though the transition is strongly first-order, there are still fluctuations present near the critical point, e.g. $D^2$ is $74\%$ of its maximal value in the VBS phase in the vicinity of the transition and the length scale of appeareance of first order signatures in the Binder cummulant of the staggered magnetization is $L \approx 4$. Mean-field theories \cite{Isaev,kotov} have not been successful for the original J-Q model. It will be interesting to see whether a reasonable description of this transition and its location can be obtained using similar techniques (e.g. mean field theory based on bond-operator formulation for $S=1/2$ introduced by Sachdev and Bhatt in Ref~\onlinecite{Subir_bo}) and this is presently being investigated. Finally, none of the finite-size signatures of first-order transitions like negative 
divergence of the Binder cummulant, crossing of energy branches etc presented in this study, have been observed in previous simulations of $SU(2)$ symmetric J-Q type models with a columnar VBS phase even for very large system sizes ($L$ up to $256$).\cite{Anders_jqlog}

\section{Acknowledgements} 

The authors would like to acknowledge useful discussions with F.~Alet, K.~Damle, R.~K.~Kaul, and R.~G.~Melko. This work is supported by the
NSF under Grant No.~DMR-0803510.


\begin{thebibliography}{999}

\bibitem{Subirbook} 
S.~Sachdev, {\it Quantum Phase Transitions} (Cambridge University Press, 1999).

\bibitem{Senthil_etal}
T. Senthil, A. Vishwanath, L. Balents, S. Sachdev, and M. P. A. Fisher, Science \textbf{303}, 1490 (2004); 
T. Senthil, L. Balents, S. Sachdev, A. Vishwanath and M. P. A. Fisher, Phys. Rev. B {\bf 70}, 144407 (2004); M. Levin and T. Senthil, Phys. Rev. B {\bf 70}, 220403 (2004).

\bibitem{read} N.~Read and S.~Sachdev, Phys. Rev. Lett. {\bf 62}, 1694 (1989). 

\bibitem{murthy} G.~Murthy and S.~Sachdev, Nucl. Phys. B {\bf 344}, 557 (1990).

\bibitem{Kuklov1} 
A.~B.~Kuklov, N. V. Prokof'ev, and B. V. Svistunov, Phys. Rev. Lett. {\bf 93}, 230402 (2004).

\bibitem{Kuklov2} 
A.~B.~Kuklov, M. Matsumoto, N. V. Prokof'ev, B. V. Svistunov, and M. Troyer, Phys. Rev. Lett. {\bf 101}, 050405 (2008).

\bibitem{motrunich}
O. I. Motrunich and A. Vishwanath, Phys. Rev. B {\bf 70}, 075104 (2004); arXiv:0805.1494.

\bibitem{takashima}
S. Takashima, I. Ichinose, and T. Matsui, Phys. Rev. B 72, 075112 (2005).

\bibitem{Anders_jq} 
A.~W.~Sandvik, Phys. Rev. Lett. {\bf 98}, 227202 (2007). 

\bibitem{Louetal} 
J.~Lou, A.~W.~Sandvik, and N.~Kawashima, Phys. Rev. B {\bf 80}, 180414(R) (2009).

\bibitem{henelius} 
P. Henelius and A. W. Sandvik, Phys. Rev. B {\bf 62}, 1102 (2000).

\bibitem{jjmodel} 
P.~Chandra and B.~Doucot, Phys. Rev. B {\bf 38}, 9335 (1988); 
A.~V.~Chubukov, Phys. Rev. B {\bf 44}, 392 (1991); 
E.~Dagotto and A.~Moreo, Phys. Rev. Lett. {\bf 63}, 2148 (1989); 
M.~P.~Gelfand, R. R. P. Singh, and D. A. Huse, Phys. Rev. B {\bf 40}, 10801 (1989).

\bibitem{melkokaul} 
R.~Melko and R.~K.~Kaul, Phys. Rev. Lett. {\bf 100}, 017203 (2008).

\bibitem{Anders_jqlog} 
A.~W.~Sandvik, Phys. Rev. Lett. {\bf 104}, 177201 (2010). 

\bibitem{kedar} 
A. Banerjee, K. Damle, and F. Alet, arXiv:1002.1375.

\bibitem{kim} 
D. H. Kim, P. A. Lee, and X.-G. Wen, Phys. Rev. Lett. {\bf 79}, 2109 (1997).


\bibitem{jiang}
F.-J. Jiang, M. Nyfeler, S. Chandrasekharan, and U.-J. Wiese, J. Stat. Mech., P02009 (2008).

\bibitem{sse} 
A. W. Sandvik, Phys. Rev. B {\bf 59}, R14157 (1999);
O.~F.~Sylju{\aa}sen and A.~W.~Sandvik, Phys. Rev. E {\bf 66}, 046701 (2002).

\bibitem{Binder} 
K.~Binder, Phys. Rev. Lett. {\bf 47}, 693 (1981); K.~Binder and D.~P.~Landau, Phys. Rev. B {\bf 30}, 1477 (1984).


\bibitem{Vollmayr}
K.~Vollmayr, J.~D.~Reger, M.~Scheucher, and K.~Binder, Z.~Phys.~B {\bf 91}, 113 (1991).

\bibitem{Cepas} O.~C\'epas, A.~P.~Young and B.~S.~Shastry, Phys. Rev. B {\bf 72}, 184408 (2005). 


\bibitem{pollock}
E. L. Pollock and D. M. Ceperley, Phys. Rev. B {\bf 36}, 8343 (1987).

\bibitem{sandvik97}
A. W. Sandvik, Phys. Rev. B {\bf 56}, 11678 (1997).



\bibitem{weak_fo} 
A.~Sen, K.~Damle, and T.~Senthil, Phys. Rev. B {\bf 76}, 235107 (2007).

\bibitem{JKmodel} 
A.~W.~Sandvik and R.~G.~Melko, Annals of Physics (N.~Y.) {\bf 321}, 1651 (2006).

\bibitem{gerber}
U. Gerber  et al., J. Stat. Mech. {\bf 2009}, P03021 (2009).


\bibitem {Fisher} 
M.~E.~Fisher and A.~N.~Berker, Phys. Rev. B {\bf 26}, 2507 (1982).

\bibitem{Lee} 
J.~Lee and J.~M.~Kosterlitz, Phys. Rev. B {\bf 43}, 3265 (1991).

\bibitem{Continentino} 
M.~A.~Continentino and A.~S.~Ferreira, Physica A {\bf 339}, 461 (2004).

\bibitem{sengupta}
P. Sengupta, A. W. Sandvik, and D. K. Campbell, Phys. Rev. B {\bf 65}, 155113 (2002).

\bibitem{troyer} 
M.~Troyer, S.~Wessel, and F.~Alet, Phys. Rev. Lett. {\bf 90}, 120201 (2003). 

\bibitem{trebst} 
S.~Trebst, D.~A.~Huse, and M.~Troyer, Phys. Rev. E {\bf 70}, 046701 (2004).


\bibitem{roger} 
R.~G.~Melko, A.~W.~Sandvik, and D.~J.~Scalapino, Phys. Rev B {\bf 69}, 100408 (R) (2004).

\bibitem{Isaev}
L. Isaev, G. Ortiz, and J. Dukelsky, J. Phys. Cond. Mat. {\bf 22}, 016006 (2009)

\bibitem{kotov}
V. N. Kotov, D.-X. Yao, A. H. Castro Neto, and D. K. Campbell,
Phys. Rev. B {\bf 80}, 174403 (2009) 

\bibitem{Subir_bo}
S.~Sachdev and R.~N.~Bhatt, Phys Rev B {\bf 41}, 9323 (1990). 

\end{thebibliography}
\end{document}